**Title: Crystallizing electrons with artificially patterned lattices**

**Author List:** Trevor G. Stanfill[1], Daniel N. Shanks[1], Michael R. Koehler[2], David G. Mandrus[3-5], Takashi Taniguchi[6], Kenji Watanabe[7], Vasili Perebeinos[8], Brian J. LeRoy[1], John R. Schaibley[1,9*]

**Affiliations:**

[1]Department of Physics, University of Arizona, Tucson, Arizona 85721, USA

[2]IAMM Diffraction Facility, Institute for Advanced Materials and Manufacturing, University of Tennessee, Knoxville, TN 37920

[3]Department of Materials Science and Engineering, University of Tennessee, Knoxville, Tennessee 37996, USA

[4]Materials Science and Technology Division, Oak Ridge National Laboratory, Oak Ridge, Tennessee 37831, USA

[5]Department of Physics and Astronomy, University of Tennessee, Knoxville, Tennessee 37996, USA

[6]Research Center for Materials Nanoarchitectonics, National Institute for Materials Science, 1-1 Namiki, Tsukuba 305-0044, Japan

[7]Research Center for Electronic and Optical Materials, National Institute for Materials Science, 1-1 Namiki, Tsukuba 305-0044, Japan

[8]Department of Electrical Engineering, University at Buffalo, Buffalo, New York 14260, USA

[9]Wyant College of Optical Sciences, The University of Arizona, Tucson, Arizona 85721, USA

*****Corresponding Author:** John Schaibley, johnschaibley@arizona.edu

**Abstract**

Wigner crystals are typically confined to ultralow temperatures where thermal motion is frozen out. Moiré superlattices in twisted two-dimensional materials have extended their stability to higher temperatures and densities, but rely on delicate stacking that fixes the lattice geometry and limits tunability. Here we demonstrate a lithographic approach that bypasses these constraints. Using high-resolution nanofabrication, we pattern a nanoscale triangular lattice directly into a graphene gate integrated with a monolayer $MoSe_2$ semiconductor. This engineered potential landscape localizes electrons into generalized Wigner crystal states that persist up to 15 K and densities of $2 \times 10^{12}$ cm$^{-2}$, representing an order of magnitude improvement over pristine monolayer $MoSe_2$. Gate-voltage control allows real-time switching between stable and unstable crystalline states, with the latter exhibiting stochastic telegraph noise from nearly degenerate configurations. This work demonstrates the ability of this platform to transform Wigner crystals from fragile, static phases into reconfigurable quantum matter.

**Introduction**

The behavior of electrons in solids can be radically reshaped by imposing a periodic superlattice potential, which modifies the band structure and can unlock entirely new quantum phases. In two-dimensional materials, such superlattices have recently been realized through twisted moiré structures, leading to remarkable discoveries—including superconductivity in magic-angle twisted bilayer graphene[1,2], the fractional quantum anomalous Hall effect[3], and generalized Wigner crystals[4-7].

Wigner crystals form when Coulomb repulsion between electrons overwhelms their kinetic energy, forcing the electrons into an ordered lattice[8]. In pristine two-dimensional systems, this fragile state survives only at ultralow density and temperatures, where thermal motion is suppressed[9,10]. Moiré superlattices have extended their stability to higher temperatures and carrier densities, but at a cost: they require painstaking alignment of atomically thin layers, and the resulting structures are fixed once assembled. This makes it difficult to design, modify, or fine-tune the underlying electronic landscape.

Here, we take a different approach. Using high-resolution lithography[11-14], we etch a nanoscale triangular lattice directly into a graphene gate, integrated with a monolayer semiconductor. This patterned gate produces a tunable potential landscape that enhances interactions, enabling the formation of generalized Wigner crystals[15] without the constraints of moiré stacking. With this method, we realize stable electron crystals at elevated temperatures and densities, and dynamically switch their configurations with gate voltages to enable access to unstable electron configurations.

**Nano-scale electronic potential**

The device design is depicted in Fig. 1a. It consists of a grounded, hBN-encapsulated MoSe$_2$ monolayer with a graphite bottom gate and a few-layer graphene (FLG) top gate. The integration of two gates allows the independent application of doping and out-of-plane electric displacement field. The top gate was nanopatterned so that doping could be spatially modulated according to the pattern geometry[16-19]. The pattern consists of a ~ 27 μm$^2$ triangular lattice of FLG holes that are 11 nm in diameter with a 40 (30) nm periodicity for device A (B). A third sample (device C) with no etched lattice was also fabricated and used as a control. Optical micrographs of devices A and B are shown in Fig. S1a,b, respectively. Additionally, to determine the regularity of the patterned lattice, atomic force microscope (AFM) topographic images and spatial Fourier transforms of the 30 nm lattice in device B were performed and are shown in Fig. S1c,d. Unless otherwise specified, the data presented in the main text were taken on device A at 1.6 K.

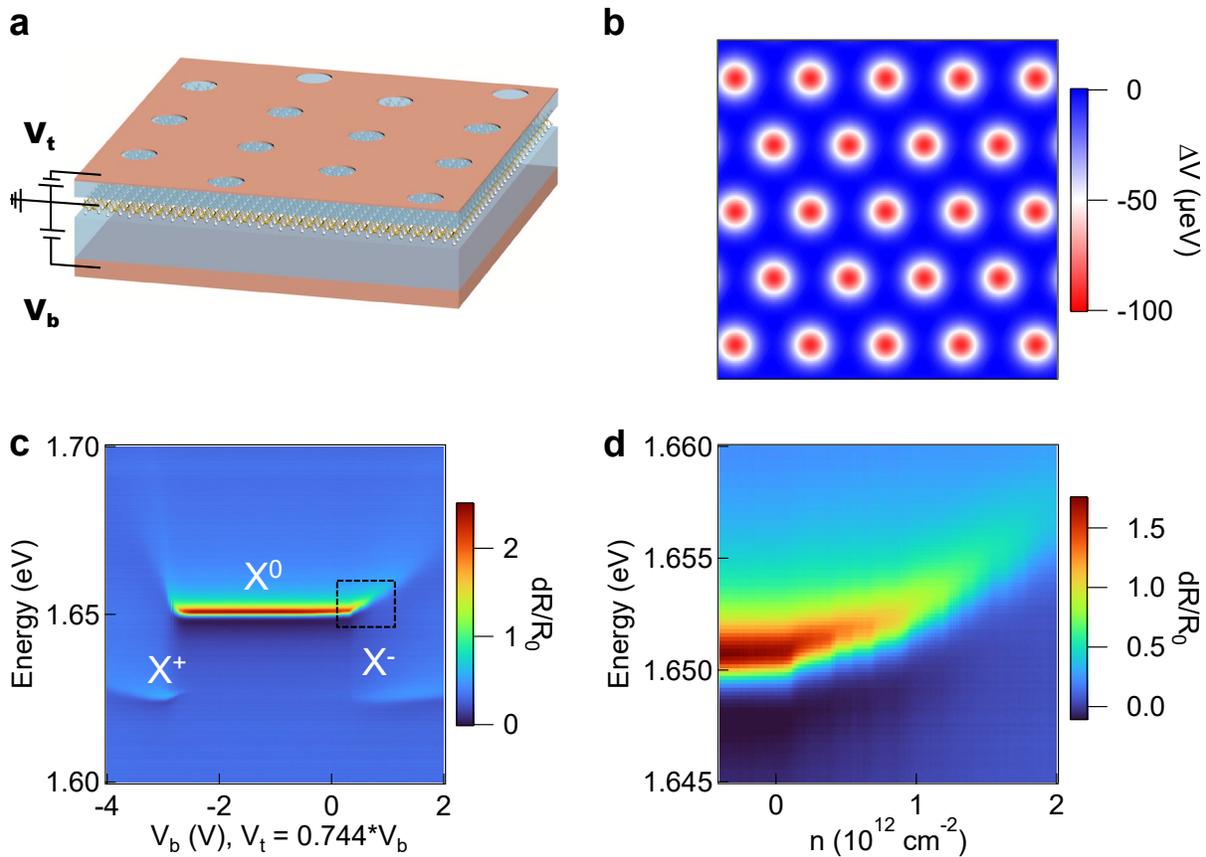

**Figure. 1. Electronic potential modulation via a nano-patterned gate. a,** Depiction of the studied heterostructure. The FLG top gate in device A (B) is patterned with a 40 (30) nm periodic triangular lattice. The MoSe$_2$ is held grounded during measurements. **b,** Simulated electric potential across the MoSe$_2$ at fixed gate voltages showing a potential variation on the order of ~ -92 μeV. Potential maxima ($\Delta V = 0$) occur away from the holes. **c,** Differential reflectivity of the sample as a function of doping, showing the neutral exciton resonance at 1.65 eV, and the X$^+$ and X$^-$ trion resonances at 1.62 eV. **d,** Differential reflectivity of the neutral exciton around the black

dashed region of **c**. The exciton exhibits discrete, step-like blue shifts and plateaus of constant energy.

When voltage was applied to the gates, a spatially varying electronic potential formed according to the geometry of the etched gate. Fig. 1b shows a continuum model simulation of the potential when the MoSe$_2$ has an electron density of $n = 1.8 \times 10^{12}$ cm$^{-2}$ and displacement field $D/\epsilon_0 = 0$ (see Methods). At this gate configuration the potential spatially varies with a depth of $\Delta V = -92$ µeV. The potential is higher away from the graphene holes than it is underneath the holes, resulting in a honeycomb of potential maxima sites. The potential energy of electrons is therefore minimized away from the patterned holes. This honeycomb potential acts to stabilize the formation of many-body electron states. Specifically, generalized Wigner crystals form in a triangular lattice, as this maximizes the interaction energy between electrons. These crystals can be pinned by either triangular sublattice of the honeycomb.

**Optical detection of insulating states**

We performed differential reflectivity measurements to investigate carrier interactions with excitons. Fig. 1c shows the reflectance of device A as a function of doping at $D/\epsilon_0 = 0$. The exciton resonance appears at 1.65 eV at low gate voltages. As doping increases, the exciton blue shifts and weakens while the positive (X$^+$) and negative (X$^-$) trion resonances appear with hole and electron doping, respectively. The neutral exciton is highlighted in Fig. 1d at the electron-doped band edge and plotted as a function of electron density (see Methods). It exhibits an overall blue shift with increased doping, with a discrete structure composed of near-vertical jumps in energy separated by plateaus of constant energy. This is not exhibited by an unpatterned sample which blue shifts smoothly[20], as shown in the gray curve in Fig. S2a,b.

To better understand the set of discrete exciton energy jumps, we measured the reflectance in an extended range of top ($V_t$) and bottom ($V_b$) gate voltages. The intensity of the exciton reflectance maxima is shown in Fig. 2a over an 8-volt range for both gates. The axes in the top right corner denote the directions of changing electric displacement field and electron doping. Consistent with Fig. 1c, the exciton intensity shows abrupt dips at the electron-doped band edge that are not observed at the hole-doped band edge. Additionally, the exciton intensity shows little dependence on applied field. Fig. 2b plots the exciton center energy at the band edge in the voltage region outlined by the dashed black box in Fig. 2a. A series of nearly-parallel lines of constant energy form close to the displacement field axis. Fig. 2c highlights the steps between energy plateaus by differentiating the center energy with respect to doping. This exciton energy derivative, $\frac{dE}{dn}$, is sensitive to the abruptness of the blue shifting and is used to determine the onset of doping in each measured sample, as shown in Fig. S2c, and to define the location of sharp blue shifts. Prominent blue shifts appear as spikes in the derivative and signal the start of an energy plateau. Fig. 2d plots a linecut of $\frac{dE}{dn}$ along the dashed line in Fig. 2c, highlighting the location of several sharp blue shifts. We distinguish peaks that are independent of displacement field with dashed lines.

Additional neighboring peaks trace features that do not occur with constant charge density and therefore cannot be attributed to Wigner crystals.

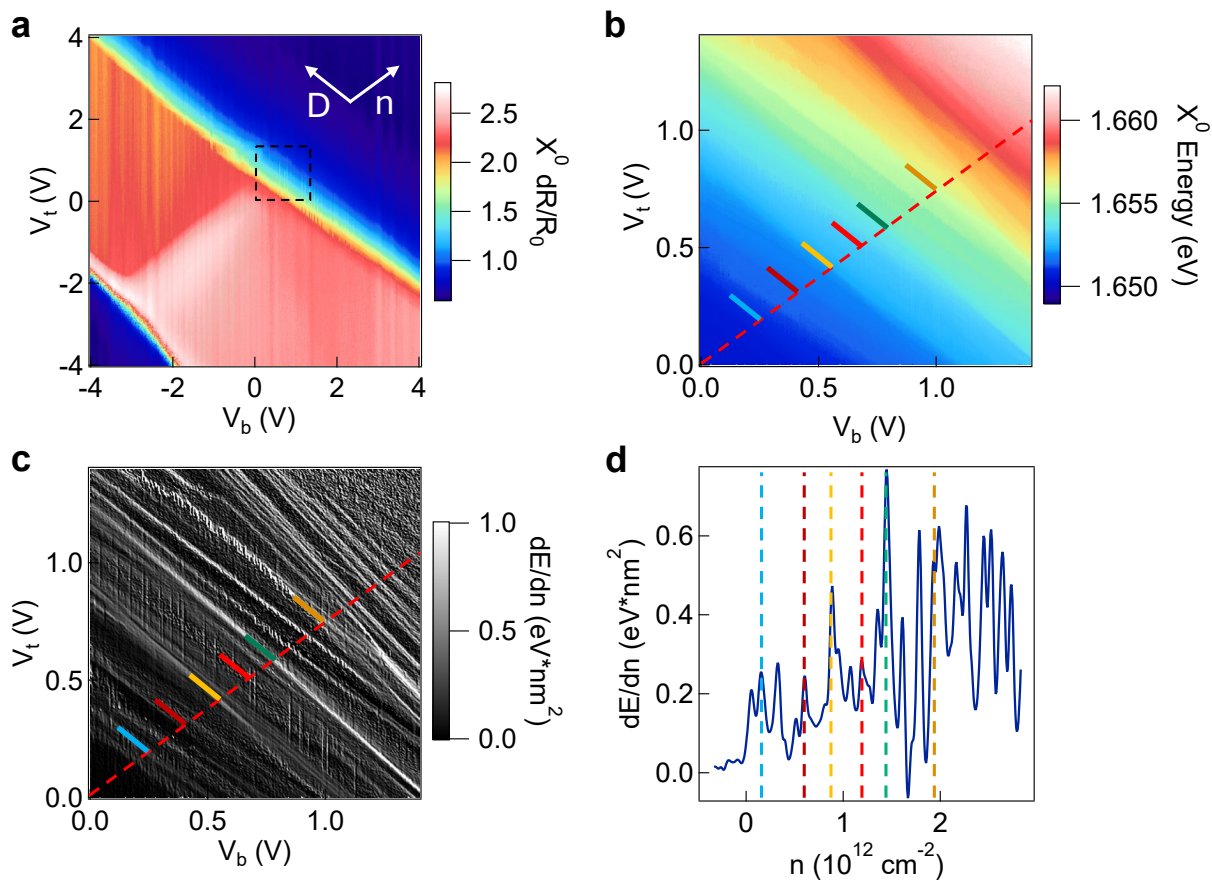

**Figure. 2. Excitonic energy jumps and plateaus near the band edge. a,** Amplitude of the exciton reflectivity as a function of $V_b$ and $V_t$. Charge neutrality is at $(V_b, V_t) = (-1.37\text{ V}, -1.07\text{ V})$, indicating the sample is inherently electron doped. The neutral exciton is replaced by $X^+$ in the bottom left and $X^-$ in the top right as hole/electron doping increase, respectively. **b,** Voltage map of the exciton energy in the black dashed box in **a**. The discrete blue shifts are arranged in bands with nearly constant charge density. Several of these are marked by colored lines. **c,** Derivative of the exciton energy with respect to doping. White lines indicate the location of sharp blue shifts. While the lines annotated are parallel to each other, a few anomalous lines are not parallel and show a stronger dependence on displacement field. **d,** A linecut of the energy derivative along the red dashed line in **(c)**, showing plateaus of constant energy separated by sharp blue shifting peaks.

**Discrete Wigner crystals**

Wigner crystallization is characterized by a dimensionless interaction parameter $r_s$. This parameter is the ratio of the Coulomb interaction energy to the electrons' kinetic energy. Quantum Monte Carlo simulations predict that, in a 2D electron gas, $r_s \geq 31$ is necessary for interactions between electrons to be large enough to allow Wigner crystal formation[21]. While this quantity is dependent on material parameters, such as the dielectric constant and effective electron mass, it is also dependent on the electron density. We note that, in monolayer MoSe₂, this criteria is met at approximately $n \approx 4 \times 10^{11}$ cm⁻² assuming an electron effective mass[22] of $m_e^* = 0.6 m_e$. In generalized Wigner crystals, however, the assisting potential can increase this critical density[5] and therefore reduce the necessary value of $r_s$.

In our measurements, we were able to resolve sharp blue shifts in $\frac{dE}{dn}$ at electron densities up to $n = 2 \times 10^{12}$ cm⁻² at $T$ = 1.6 K. We attribute these sharp blue shifts in the neutral exciton energy to transitions between insulating generalized Wigner crystals that stabilize around the potential at the beginning of a plateau and unpin from the potential at the end of the plateau. After forming, the crystal opens a band gap, such that further doping serves to raise the Fermi energy and eventually unpin the crystal. Once the crystal is unpinned, further doping serves to increase the electron density and raise the exciton energy[23].

As the electron density increases, the size of an emergent Wigner crystal will pass through geometrically commensurate multiples of the triangular potential induced by the patterned gate. Specifically, we consider when $Q \times \ell_{\text{Wigner}} = P \times \ell_{\text{pattern}}$, where $Q$ are positive integers, $P$ are positive integers and irrational values determined by the patterned lattice, and $\ell_{\text{Wigner/pattern}}$ are the lattice constants of the Wigner crystal and patterned gate, respectively. This condition can alternatively be written as $\frac{Q^2}{P^2} = \frac{n_{\text{Wigner}}}{n_{\text{pattern}}}$, where $Q^2$ intuitively describes the number of electrons contained in $P^2$ number of patterned lattice unit cells. For electron densities satisfying this condition, the crystal can be oriented so that none of the electrons reside on potential minima sites. Several commensurate crystals satisfying this condition are shown in Fig. 3a with $Q$ values ranging from 1 to 3.

Generalized Wigner crystals are distinguished from other correlated charge states, such as Mott insulators[24], by their existence at well-known[6] fractional fillings of a superlattice. Specifically, because Wigner crystals only occur with a triangular geometry, non-triangular states exhibiting correlated behavior in moiré systems, such as the $\nu = \frac{2}{3}$ filled state, require a potential strong enough to trap individual electrons and do not satisfy the commensuration condition. The relatively weak trapping potential induced by the patterned gate only stabilizes configurations preferred by Coulomb repulsion.

To validate this interpretation, we studied the location of the $\frac{dE}{dn}$ peaks as a function of the Fermi wave vector, $k_f = \sqrt{\pi n}$, plotted as the blue curve in Fig. 3b. Along this x-axis, crystals form periodically with spacing determined by the value of $P$. This is illustrated in Fig. 3b by the series of red, yellow, and blue lines that denote the $P = 1$, $P = \sqrt{3}$, and $P = 2$ series of crystals, respectively. The black curve shows a simulated series of Gaussian peaks that represent an exciton energy derivative dominated by Wigner crystal induced energy spikes. This was simulated at each commensurate value of $k_f$ up to $P = 3\sqrt{3}$ (see Methods). The Fourier transform of $\frac{dE}{dn}$ v.s. $k_f$ reveals the periodic structure which arises from peaks repeating at integer $Q$ values. Fig. 3c shows the Fourier transform of the $\frac{dE}{dn}$ curves for the 40 nm (blue), the 30 nm (teal), the simulated data (black), and an un-patterned control sample (grey). The x-axis for each of the Fourier transforms is divided by the periodic lattice spacing to allow them to all be plotted in terms of $P$-series. The un-patterned control device was plotted in terms of $P$ by assuming it had a 40 nm lattice and dividing accordingly. Peaks should occur at integer multiples of 1, $\sqrt{3}$, and $\sqrt{7}$ for the range of $P$ values shown. These series of commensurate crystals are again marked by vertical lines, which are in excellent agreement with the 40 nm and 30 nm patterned sample peaks.

To assess consistency of the location of Fourier periodic peaks, Fourier transforms of $\frac{dE}{dn}$ curves on device A during different cooldown cycles and at different probing locations across the sample are shown in Fig. S2d. The location of peaks is consistent, providing evidence that the Fourier analysis is a robust way to analyze the blue shifts. While it is tempting to identify individual peaks and associate them with specific commensurate crystals, the presence of field-dependent peaks with a non-Wigner crystal origin makes it impossible to distinguish between the two. Specific crystals can only be attributed using data containing doping and field dependence as shown in Fig. S3a,b. The evolution of these features with temperature is studied in Fig. S3c-f showing that all features are gone by T = 45 K.

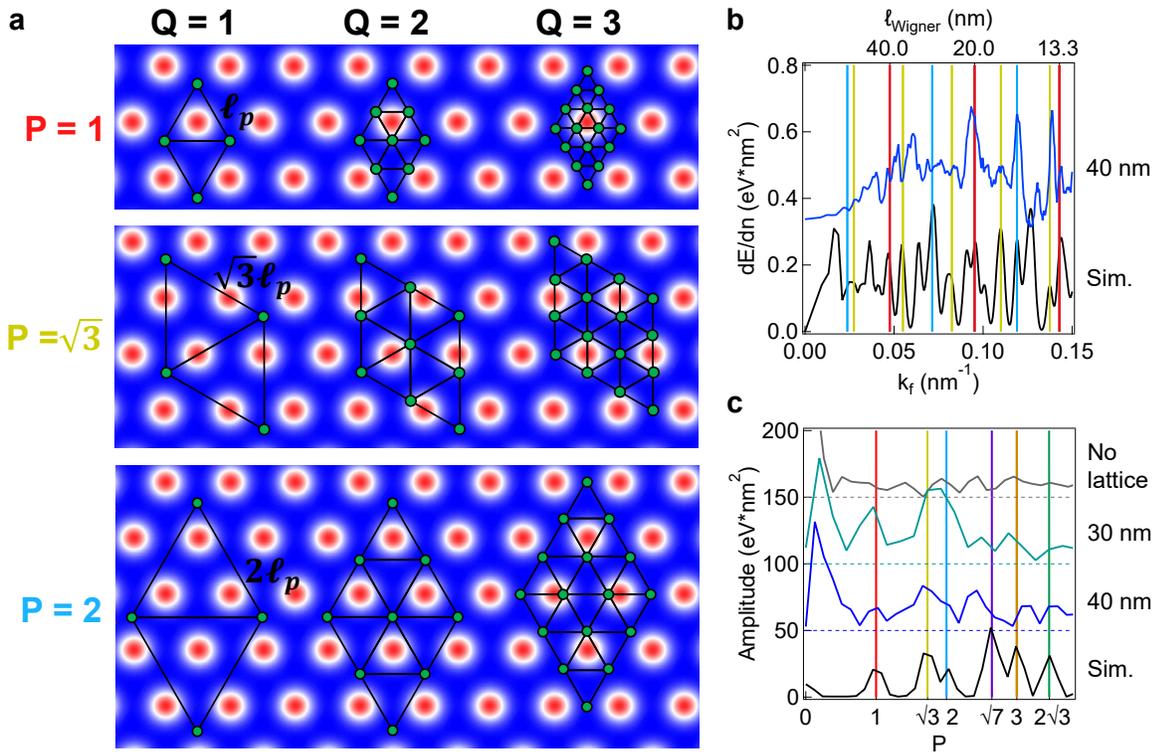

**Figure. 3**. $k_f$-**Periodic Wigner crystals commensurate with the patterned lattice. a,** Illustration of the orientation of stable Wigner crystals, with electrons depicted as green dots. The top row demonstrates crystals that satisfy $Q\ell_{Wigner} = P\ell_{pattern}$ with $P = 1$. As $Q$ increases, the crystal density increases. The following rows depict rotated $P = \sqrt{3}$ crystals and $P = 2$ crystals. The commensurate inter-electron spacing is set by the distance between potential maxima sites and is labelled for the crystals in the first column. **b,** Exciton energy derivative with respect to doping, and simulated commensurate Wigner crystals plotted as functions of the Fermi wave vector. Along this axis Wigner crystals with fixed values of $P$ occur periodically. The $P = 1$, $P = \sqrt{3}$, and $P = 2$ series are depicted by red, yellow, and blue lines, respectively. **c,** Fourier transformation of the curves in **b**, as well as the other fabricated samples, with respect to $k_f$. The simulated curve is plotted in black and shows peaks at the commensurate values of $P$. Device A (blue curve) shows excellent agreement with the simulated curve. Device B (teal curve) shows qualitatively similar peaks to device A. Device C (grey curve) is dominated by aperiodic noise.

In contrast to the peaks observed in the patterned samples, the unpatterned sample consists of aperiodic noise. This is highlighted in Fig. S4a where the Fourier transforms of the three samples are plotted over an extended range of $P$ values. For the unpatterned sample the noise has a constant amplitude for all values of $P$ shown. Temperature dependence of the $\frac{dE}{dn}$ peaks in device A and their Fourier transform are shown in Fig. S4b,c respectively. The $\frac{dE}{dn}$ and Fourier peaks diminish with temperature, signifying crystal melting, and disappear at $T = 30$ K. Interestingly, only the $P = 1$ and $P = 2$ series of crystals persist to $T = 15$ K. Despite the increase in signal-to-noise ratio gained by using Fourier analysis to study periodic features in our $\frac{dE}{dn}$ curves, structure was only observed at the hole-doped band edge of device A (Fig. S4). Fig. S4 d and e show the exciton blue shift curve and the Fourier transformation of the associated $\frac{dE}{dn}$ curve respectively. The blue shift is dominated by noise that could be caused by an extrinsic effect and is left as the subject of future study.

**Field-induced telegraph noise**

Although the exciton energy was generally stable over time for fixed gate voltages, we discovered that with the application of a fixed displacement field, a stochastic fluctuation in the exciton energy was observed between $n = 7 \times 10^{11}$ cm$^{-2}$ and $n = 1.7 \times 10^{12}$ cm$^{-2}$. Fig. 4a shows the exciton energy as a function of doping when a $D/\epsilon_0 = 0.35 \; \frac{\text{V}}{\text{nm}}$ displacement field was applied. In addition to the previously observed step-like blue shifts, we observed energy fluctuations on the order of ~ 250 μeV. These fluctuations are highlighted in the inset, where point-to-point variations in energy occur on top of the blue shifts. The red, green, and blue dots in the inset correspond to the curves in Fig. 4b where we measured the exciton energy as a function of time while keeping the doping and displacement field (sample voltages) fixed. The energy was measured with a temporal resolution of 100 msec and switches between two stable values.

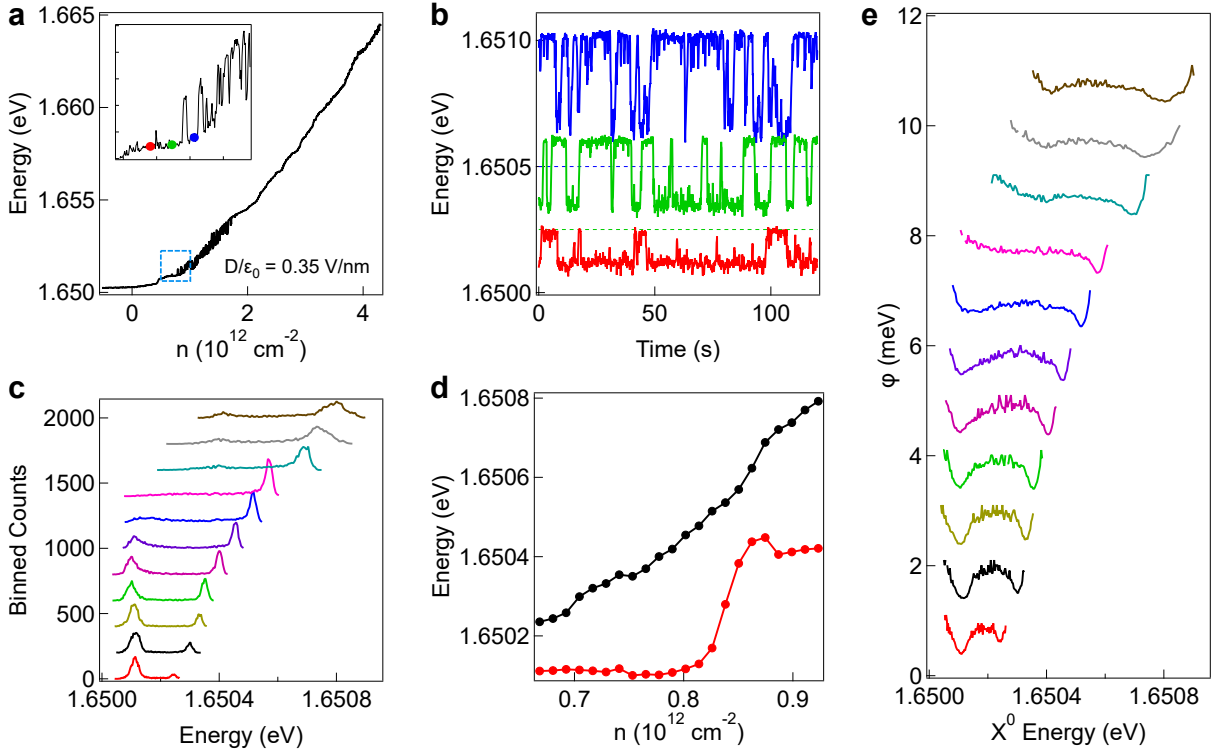

**Figure. 4. Quantum telegraph noise arising from Wigner crystal switching. a,** Exciton differential reflectance energy as a function of doping when a fixed displacement field is applied. Stochastic jumps in the energy occur around $n = 1 \times 10^{12}$ cm$^{-2}$. The inset shows the energy fluctuations from one point to the next on the order of $\sim 250$ μeV. **b,** Exciton energy variation with time at the densities highlighted by the colored dots in the inset of **a**. The energy fluctuates between two stable values with switching times limited by our 100 msec temporal resolution. The ΔE between the stable values becomes larger with doping. **c,** Histograms of the energy as a function of time. The curves increase in charge density from $n = 6.8 \times 10^{11}$ cm$^{-2}$ to $n = 9.23 \times 10^{11}$ cm$^{-2}$ in $\Delta n = 2.43 \times 10^{10}$ cm$^{-2}$ increments. The exciton is mostly in the low energy state at $n = 6.8 \times 10^{11}$ cm$^{-2}$ and spends more time in the higher energy state as doping increases, until it is purely in the high energy state at the pink curve. A new low energy state starts to emerge at larger densities. **d,** Energy of the two states as a function of doping, extracted through Lorentzian fits to the 2-peak histograms. The lower energy state is composed of plateaus with sharp blue shifts that move from one plateau to the next, while the high energy state exhibits a smoothly varying blue shift. **e,** Calculated potential energy of the system for each curve in **c**.

This effect occurs only at low temperatures where Wigner crystal features are well-resolved, and is present across the entire sample at specific densities and fields. Additionally, because this is a far-field optical measurement with a probing area of $\sim 1275$ patterned holes, this is attributed to a collective electronic effect. This behavior is characteristic of quantum telegraph noise[25-27], where two nearly-degenerate quantum states show stochastic state-switching. This type of noise is expected but previously unreported in TMD Wigner crystal systems undergoing pinning and depinning processes[28], which would not be expected for other correlated insulating states[29, 30]. The probability of finding the exciton energy in either state was determined by plotting binned histograms of the energy over many switching cycles. In Fig. 4c, histograms of 10-minute scans are plotted for increasing charge densities. The histograms show that the energy fluctuates between two stable states with Lorentzian noise distributed around each state. As the density increases, the proportion of time spent in the higher energy state increases until the energy is nearly stable at the high energy state. This is followed at even higher doping by the emergence of a second low energy peak. Fig. 4d tracks how the energy of each state evolves with doping and shows the high energy peak (black points) blue shifts with doping while the low energy peak (red) discretely transitions between two constant values. The low energy state is characteristic of a pinned Wigner crystal switching between two stable configurations whereas the high energy state is indicative of an uncorrelated, free electronic state. We note that this behavior is only observed above the maximum Wigner crystal density in pristine monolayer $MoSe_2$ where the electrons are normally expected to be uncorrelated, providing further evidence that the potential stabilizes Wigner crystals to elevated charge densities.

Using the Boltzmann factor, $P(\varepsilon) = e^{\frac{-\phi(\varepsilon)}{kT}}$, a landscape of the potential $\phi$ for the two-state system was calculated to determine their energy separation (Figure 4e). The potential shows two valleys separated by a potential barrier with valley minima that differ by as much as ~ 200 µeV. The shape of the potential is consistent with reports of random telegraph noise in magnetic tunnel junctions[25] and the barrier height indicates the two states are nearly degenerate at this doping and field. This effect was surprising and illustrates the opportunities of studying generalized Wigner crystal systems with an imprinted periodic potential. At densities that nearly satisfy the commensuration conditions, the pinned Wigner crystal and uncorrelated free electron configurations are nearly degenerate and give rise to dynamic reconfiguration that cannot be observed in unassisted Wigner crystal systems.

**Discussion**

We have demonstrated a new pathway to create and control correlated quantum phases—without the tedious layer-by-layer assembly of moiré heterostructures—by directly etching nanoscale patterns into gate electrodes. This approach delivers tunable superlattice potentials whose periodicity and symmetry are not limited by the underlying atomic crystal. While here we employed a triangular lattice to stabilize generalized Wigner crystals, our technique is completely general: square lattices, Kagome geometries, quasicrystals, or anisotropic patterns are all possible. By tailoring the landscape of the trapping potential, we can not only pin known electron crystals but also create conditions for entirely new phases that have no counterpart in natural materials.

This ability to "write" custom electronic landscapes on demand transforms 2D materials into programmable quantum matter platforms. It opens the door to exploring unconventional correlated states, exotic excitations, and artificial electronic metamaterials that lie far beyond the reach of moiré engineering. In short, we have turned patterned gates into a quantum design tool—one that could reshape how we think about building and discovering new states of matter.

# References

(1) Cao, Y.; Fatemi, V.; Fang, S.; Watanabe, K.; Taniguchi, T.; Kaxiras, E.; Jarillo-Herrero, P. Unconventional Superconductivity in Magic-Angle Graphene Superlattices. *Nature* **2018**, *556* (7699), 43–50. DOI: 10.1038/nature26160.
(2) Bistritzer, R.; MacDonald, A. H. Moiré Bands in Twisted Double-Layer Graphene. *Proceedings of the National Academy of Sciences* **2011**, *108* (30), 12233–12237. DOI: 10.1073/pnas.1108174108 (acccessed 2025/08/07).
(3) Cai, J.; Anderson, E.; Wang, C.; Zhang, X.; Liu, X.; Holtzmann, W.; Zhang, Y.; Fan, F.; Taniguchi, T.; Watanabe, K.; et al. Signatures of Fractional Quantum Anomalous Hall States in Twisted $MoTe_2$. *Nature* **2023**, *622* (7981), 63–68. DOI: 10.1038/s41586-023-06289-w.
(4) Kiper, N.; Adlong, H. S.; Christianen, A.; Kroner, M.; Watanabe, K.; Taniguchi, T.; İmamoğlu, A. Confined Trions and Mott-Wigner States in a Purely Electrostatic Moiré Potential. *Physical Review X* **2025**, *15* (1), 011049. DOI: 10.1103/PhysRevX.15.011049.
(5) Regan, E. C.; Wang, D.; Jin, C.; Bakti Utama, M. I.; Gao, B.; Wei, X.; Zhao, S.; Zhao, W.; Zhang, Z.; Yumigeta, K.; et al. Mott and Generalized Wigner Crystal States in $WSe_2/WS_2$ Moiré Superlattices. *Nature* **2020**, *579* (7799), 359–363. DOI: 10.1038/s41586-020-2092-4.
(6) Xu, Y.; Liu, S.; Rhodes, D. A.; Watanabe, K.; Taniguchi, T.; Hone, J.; Elser, V.; Mak, K. F.; Shan, J. Correlated Insulating States at Fractional Fillings of Moiré Superlattices. *Nature* **2020**, *587* (7833), 214–218. DOI: 10.1038/s41586-020-2868-6.
(7) Li, H.; Li, S.; Regan, E. C.; Wang, D.; Zhao, W.; Kahn, S.; Yumigeta, K.; Blei, M.; Taniguchi, T.; Watanabe, K.; et al. Imaging Two-Dimensional Generalized Wigner Crystals. *Nature* **2021**, *597* (7878), 650–654. DOI: 10.1038/s41586-021-03874-9.
(8) Wigner, E. On the Interaction of Electrons in Metals. *Physical Review* **1934**, *46* (11), 1002–1011. DOI: 10.1103/PhysRev.46.1002.
(9) Zhou, Y.; Sung, J.; Brutschea, E.; Esterlis, I.; Wang, Y.; Scuri, G.; Gelly, R. J.; Heo, H.; Taniguchi, T.; Watanabe, K.; et al. Bilayer Wigner Crystals in a Transition Metal Dichalcogenide Heterostructure. *Nature* **2021**, *595* (7865), 48–52. DOI: 10.1038/s41586-021-03560-w.
(10) Smoleński, T.; Dolgirev, P. E.; Kuhlenkamp, C.; Popert, A.; Shimazaki, Y.; Back, P.; Lu, X.; Kroner, M.; Watanabe, K.; Taniguchi, T.; et al. Signatures of Wigner Crystal of Electrons in a Monolayer Semiconductor. *Nature* **2021**, *595* (7865), 53–57. DOI: 10.1038/s41586-021-03590-4.
(11) Shanks, D. N.; Mahdikhanysarvejahany, F.; Koehler, M. R.; Mandrus, D. G.; Taniguchi, T.; Watanabe, K.; LeRoy, B. J.; Schaibley, J. R. Single-Exciton Trapping in an Electrostatically Defined Two-Dimensional Semiconductor Quantum Dot. *Physical Review B* **2022**, *106* (20), L201401. DOI: 10.1103/PhysRevB.106.L201401.
(12) Shanks, D. N.; Mahdikhanysarvejahany, F.; Muccianti, C.; Alfrey, A.; Koehler, M. R.; Mandrus, D. G.; Taniguchi, T.; Watanabe, K.; Yu, H.; LeRoy, B. J.; et al. Nanoscale Trapping of Interlayer Excitons in a 2D Semiconductor Heterostructure. *Nano Letters* **2021**, *21* (13), 5641–5647. DOI: 10.1021/acs.nanolett.1c01215.
(13) Forsythe, C.; Zhou, X.; Watanabe, K.; Taniguchi, T.; Pasupathy, A.; Moon, P.; Koshino, M.; Kim, P.; Dean, C. R. Band Structure Engineering of 2D Materials Using Patterned Dielectric Superlattices. *Nature Nanotechnology* **2018**, *13* (7), 566–571. DOI: 10.1038/s41565-018-0138-7.
(14) Shanks, D. N.; Mahdikhanysarvejahany, F.; Stanfill, T. G.; Koehler, M. R.; Mandrus, D. G.; Taniguchi, T.; Watanabe, K.; LeRoy, B. J.; Schaibley, J. R. Interlayer Exciton Diode and Transistor. *Nano Letters* **2022**, *22* (16), 6599–6605. DOI: 10.1021/acs.nanolett.2c01905.
(15) Hubbard, J. Generalized Wigner Lattices in One Dimension and Some Applications to Tetracyanoquinodimethane (TCNQ) Salts. *Physical Review B* **1978**, *17* (2), 494–505. DOI: 10.1103/PhysRevB.17.494.
(16) Thureja, D.; Imamoglu, A.; Smoleński, T.; Amelio, I.; Popert, A.; Chervy, T.; Lu, X.; Liu, S.; Barmak, K.; Watanabe, K.; et al. Electrically Tunable Quantum Confinement of Neutral Excitons. *Nature* **2022**, *606* (7913), 298–304. DOI: 10.1038/s41586-022-04634-z.
(17) Hu, J.; Lorchat, E.; Chen, X.; Watanabe, K.; Taniguchi, T.; Heinz, T. F.; Murthy, P. A.; Chervy, T. Quantum Control of Exciton Wave Functions in 2D Semiconductors. *Science Advances* **2024**, *10* (12), eadk6369. DOI: doi:10.1126/sciadv.adk6369.
(18) Moshchalkov, V. V.; Baert, M.; Metlushko, V. V.; Rosseel, E.; Van Bael, M. J.; Temst, K.; Bruynseraede, Y.; Jonckheere, R. Pinning by an Antidot Lattice: The Problem of the Optimum Antidot Size. *Physical Review B* **1998**, *57* (6), 3615–3622. DOI: 10.1103/PhysRevB.57.3615.
(19) Pedersen, T. G.; Flindt, C.; Pedersen, J.; Jauho, A.-P.; Mortensen, N. A.; Pedersen, K. Optical Properties of Graphene Antidot Lattices. *Physical Review B* **2008**, *77* (24), 245431. DOI: 10.1103/PhysRevB.77.245431.
(20) Back, P.; Zeytinoglu, S.; Ijaz, A.; Kroner, M.; Imamoğlu, A. Realization of an Electrically Tunable Narrow-Bandwidth Atomically Thin Mirror Using Monolayer MoSe$_2$. *Physical Review Letters* **2018**, *120* (3), 037401. DOI: 10.1103/PhysRevLett.120.037401.
(21) Drummond, N. D.; Needs, R. J. Phase Diagram of the Low-Density Two-Dimensional Homogeneous Electron Gas. *Physical Review Letters* **2009**, *102* (12), 126402. DOI: 10.1103/PhysRevLett.102.126402.
(22) Kormányos, A.; Burkard, G.; Gmitra, M.; Fabian, J.; Zólyomi, V.; Drummond, N. D.; Fal'ko, V. k·p Theory for Two-Dimensional Transition Metal Dichalcogenide Semiconductors. *2D Materials* **2015**, *2* (2), 022001. DOI: 10.1088/2053-1583/2/2/022001.
(23) Van Tuan, D.; Dery, H. Excitons in Periodic Potentials. *Physical Review B* **2023**, *108* (8), L081301. DOI: 10.1103/PhysRevB.108.L081301.
(24) Padhi, B.; Chitra, R.; Phillips, P. W. Generalized Wigner Crystallization in Moiré Materials. *Physical Review B* **2021**, *103* (12), 125146. DOI: 10.1103/PhysRevB.103.125146.
(25) Hayakawa, K.; Kanai, S.; Funatsu, T.; Igarashi, J.; Jinnai, B.; Borders, W. A.; Ohno, H.; Fukami, S. Nanosecond Random Telegraph Noise in in-Plane Magnetic Tunnel Junctions. *Physical Review Letters* **2021**, *126* (11), 117202. DOI: 10.1103/PhysRevLett.126.117202.
(26) Pistol, M. E.; Castrillo, P.; Hessman, D.; Prieto, J. A.; Samuelson, L. Random Telegraph Noise in Photoluminescence from Individual Self-Assembled Quantum Dots. *Physical Review B* **1999**, *59* (16), 10725–10729. DOI: 10.1103/PhysRevB.59.10725.
(27) Ralls, K. S.; Skocpol, W. J.; Jackel, L. D.; Howard, R. E.; Fetter, L. A.; Epworth, R. W.; Tennant, D. M. Discrete Resistance Switching in Submicrometer Silicon Inversion Layers: Individual Interface Traps and Low-Frequency (1/f?) Noise. *Physical Review Letters* **1984**, *52* (3), 228–231. DOI: 10.1103/PhysRevLett.52.228.

(28) Reichhardt, C.; Reichhardt, C. J. O. Depinning, Melting, and Sliding of Generalized Wigner Crystals in Moiré Systems. *Physical Review Research* **2025**, *7* (1), 013155. DOI: 10.1103/PhysRevResearch.7.013155.
(29) Grüner, G. The Dynamics of Charge-Density Waves. *Reviews of Modern Physics* **1988**, *60* (4), 1129–1181. DOI: 10.1103/RevModPhys.60.1129.
(30) Reichhardt, C.; Reichhardt, C. J. O. Noise and Thermal Depinning of Wigner Crystals. *Journal of Physics: Condensed Matter* **2023**, *35* (32), 325603. DOI: 10.1088/1361-648X/acd218.
(31) Geick, R.; Perry, C. H.; Rupprecht, G. Normal Modes in Hexagonal Boron Nitride. *Physical Review* **1966**, *146* (2), 543–547. DOI: 10.1103/PhysRev.146.543.

**Methods**

**Sample fabrication**

The samples studied in this work were fabricated layer-by-layer using the dry transfer polymer method. Each 2D material was first exfoliated onto a $SiO_2$ / Si substrate and found under an optical microscope. Material thickness was then determined by optical contrast and by AFM topography. Device A (40 nm) had top and bottom hBN layers with thicknesses of $l_t = 14.5$ nm and $l_b = 19.5$ nm, respectively. Additionally, device B (30 nm) had hBN layers with thicknesses of $l_t = 24.5$ nm and $l_b = 10.4$ nm and device C (no lattice) had a top hBN layer with thickness $l_t = 16.3$ nm. The layers were then stacked and placed on a substrate with a rectangular grid of gold markers. After transferring, the lattices were patterned using electron beam lithography into a 70 nm thick layer of 950 K A2 poly (methyl methacrylate), PMMA, that sits over the top FLG gate. The sample was then developed and the exposed FLG was etched using an $O_2$ reactive ion etch plasma for 6 seconds with 50 sccm of gas at 100 mTorr. Chromium and gold contacts were then appended to the device using electron beam lithography and thermal metal evaporation to achieve thicknesses of 5 nm of chromium and 35 nm of gold.

**Optical measurements**

Optical measurements were performed in an AttoDRY 2100 cryostat at a base temperature of $T = 1.6$ K. Differential reflectivity was performed with an NKT Photonics supercontinuum white-light laser with a 78 MHz repetition rate at a laser power between 70 and 800 nW. A 650 nm long-pass and an 800 nm short-pass were used to filter the light. A laser spot size of ~1 μm was used to probe the sample. In device A, this spot size covered approximately ~ 500 lattice sites. The differential reflectivity was calculated as $\frac{dR}{R_0} = \frac{R-R_0}{R_0}$ where $R, R_0$ were defined as the background corrected sample reflectivity and substrate reflectivity. The intensity reported in Fig. 2a was calculated by finding the maximum $\frac{dR}{R_0}$ value over a restricted energy range only containing the neutral exciton. The exciton energy was extracted through a Fano line-shape fit to the exciton peak.

**Simulation of the electronic potential**

The interaction parameter $r_s$ is the ratio of kinetic energy to Coulomb energy and is calculated as $r_s = \frac{\sqrt{\pi} m_e^* e^2}{\epsilon_0 \epsilon h^2 \sqrt{n}}$ where $m_e^*$ is the electron effective mass, $e$ is the charge of the electron, $h$ is Planck's constant, $\epsilon$ is the dielectric constant, and $n$ is the electron density.

We employ the Thomas-Fermi model to simulate charge carrier density in the MoSe$_2$ layer. The Poisson equation is used for the self-consistent potential $V(x, y, z)$:

$$\epsilon_0 \epsilon(x, y, z) \left( \frac{d^2 V(x,y,z)}{dx^2} + \frac{d^2 V(x,y,z)}{dy^2} + \frac{d^2 V(x,y,z)}{dz^2} \right) = -\rho(x, y, z), \tag{1}$$

where $\epsilon(x, y, z)$ is the dielectric constant of either h-BN[31], $\epsilon_{hBN} = 5.06$, or air, $\epsilon_{air} = 1.0$. The carrier density, $\rho(x, y, z)$, is given by:

$$\rho(x, y, z) = n(x, y) \exp(-z^2/\sigma^2)/(\sqrt{\pi}\sigma), \tag{2}$$

where $n(x, y)$ is 2D carrier density on the MoSe$_2$ layer and $\sigma = 4$ Å is used to mimic the finite width of the electron cloud. The 2D carrier density $n(x, y)$ depends on the electrostatic potential on the MoSe$_2$ layer at $V(x, y, z = 0)$, which is found self-consistently via

$$n(V) = k_B T \cdot \frac{2m_e^*}{\pi \hbar^2} \cdot \ln\left( \frac{1+e^{(-eV-E_g/2)/k_B T}}{1+e^{(eV-E_g/2)/k_B T}} \right), \tag{3}$$

where $m_e^* = 0.6 m_e$ in units of electron mass $m_e$, $E_g = 2.5$ eV is the experimentally determined electrical bandgap of the semiconductor, and $k_B$ is the Boltzmann constant.

**Simulation of $k_f$-periodic peaks**

A Gaussian with a fixed width and amplitude was first defined by fitting a Gaussian to an isolated, low-amplitude peak in device A as a function of charge density. The series of these peaks that satisfy the commensuration equation for $P = 1, 2, 3$, $P = \sqrt{7}$, and $P = \sqrt{3}, 2\sqrt{3}, 3\sqrt{3}$ were then superimposed and plotted along $k_f$. The simulated curve contained the same number of data points as the data from device A to accurately simulate the peak structure. An attempt was made to match individual $\frac{dE}{dn}$ peaks between devices A and B, however, due to the large number of peaks in device B we were unable to find a statistically significant correlation between peaks in the two samples without using Fourier analysis.

**Fourier analysis of patterned lattice**

A 2D spatial Fourier transformation was calculated over a 500 nm × 500 nm region of the AFM topographic image in Fig. S1c. The lattice constant was extracted along the k-space lattice vectors by fitting 2D Gaussian functions to the individual peaks and calculating the lattice constant from the center of each Gaussian. These were then averaged to give a lattice constant of $a = 28.67$ nm.

The width of the Gaussian fits was then used to determine the regularity of the lattice constant to be $\Delta a = 2.65$ nm.

## Acknowledgements

We thank Allan H. MacDonald, Huiyuan Zheng, and Hanan Dery for fruitful discussions.

**Funding Acknowledgement:** JRS and BJL acknowledge support from NSF Grant Nos. ECCS-2054572, ECCS-2428575, and AFOSR Grant No. FA9550-22-1-0220. JRS and VP acknowledge support from AFOSR Grant No. FA9550-22-1-0312. JRS acknowledges support from AFOSR Grant No. FA9550-22-1-0113. JRS acknowledges support from the 2023 Technology Research Initiative Fund - National Security Systems - Novel Materials Project, administered by the University of Arizona Office of Research and Partnerships (ORP), funded under Proposition 301, the Arizona Sales Tax for Education Act, in 2000. This work was supported by the Gordon and Betty Moore Foundation, grant DOI 10.37807/GBMF13840. BJL acknowledges support from NSF Grant No. ECCS-2122462. VP acknowledges support from NSF Grant No. ECCS-2235276. KW and TT acknowledge support from the JSPS KAKENHI (Grant Numbers 21H05233 and 23H02052), the CREST (JPMJCR24A5), JST and World Premier International Research Center Initiative (WPI), MEXT, Japan. DGM acknowledges support from the Gordon and Betty Moore Foundation's EPiQS Initiative grant GBMF9069. The Plasmatherm reactive ion etcher used in this study was acquired through an NSF MRI grant, Award No. ECCS-1725571.

## Author contributions

JRS and BJL supervised the research. TS, DNS, JRS and BJL conceived of the project. MRK and DGM grew and characterized the MoSe$_2$ bulk crystals. TT and KW grew and characterized the bulk hBN crystals. VP calculated the charge density and electronic potential in the semiconductor layer. TS fabricated the devices and performed all of the experiments. TS analyzed the data and interpreted the results with guidance from JRS and BJL. TS wrote the manuscript with input from JRS and BJL.

**Supporting Information for: Crystallizing electrons with artificially patterned lattices**

**Author List:** Trevor G. Stanfill[1], Daniel N. Shanks[1], Michael R. Koehler[2], David G. Mandrus[3-5], Takashi Taniguchi[6], Kenji Watanabe[7], Vasili Perebeinos[8], Brian J. LeRoy[1], John R. Schaibley[1,9*]

**Affiliations:**

[1]Department of Physics, University of Arizona, Tucson, Arizona 85721, USA

[2]IAMM Diffraction Facility, Institute for Advanced Materials and Manufacturing, University of Tennessee, Knoxville, TN 37920

[3]Department of Materials Science and Engineering, University of Tennessee, Knoxville, Tennessee 37996, USA

[4]Materials Science and Technology Division, Oak Ridge National Laboratory, Oak Ridge, Tennessee 37831, USA

[5]Department of Physics and Astronomy, University of Tennessee, Knoxville, Tennessee 37996, USA

[6]Research Center for Materials Nanoarchitectonics, National Institute for Materials Science, 1-1 Namiki, Tsukuba 305-0044, Japan

[7]Research Center for Electronic and Optical Materials, National Institute for Materials Science, 1-1 Namiki, Tsukuba 305-0044, Japan

[8]Department of Electrical Engineering, University at Buffalo, Buffalo, New York 14260, USA

[9]Wyant College of Optical Sciences, The University of Arizona, Tucson, Arizona 85721, USA

*****Corresponding Author:** John Schaibley, johnschaibley@arizona.edu

**Table of contents**

## 1. Device images and lattice regularity

Images of the 40 nm and 30 nm patterned samples are shown in Fig. S1 a and b, respectively. The region of the top gate that was patterned is outlined in blue and extends over a roughly 28 μm² area. AFM characterization of the 30 nm triangular lattice is plotted in Fig. S1c and shows the lattice patterned in PMMA over a 750 nm × 750 nm square area. A spatial Fourier transformation of the 30 nm lattice is taken of the AFM topography in Fig. S1d to determine the lattice regularity. By fitting 2D gaussian functions to each peak, the lattice constant is determined to be $a = 28.67$ nm and the full width at half maximum is $\Delta a = 2.65$ nm.

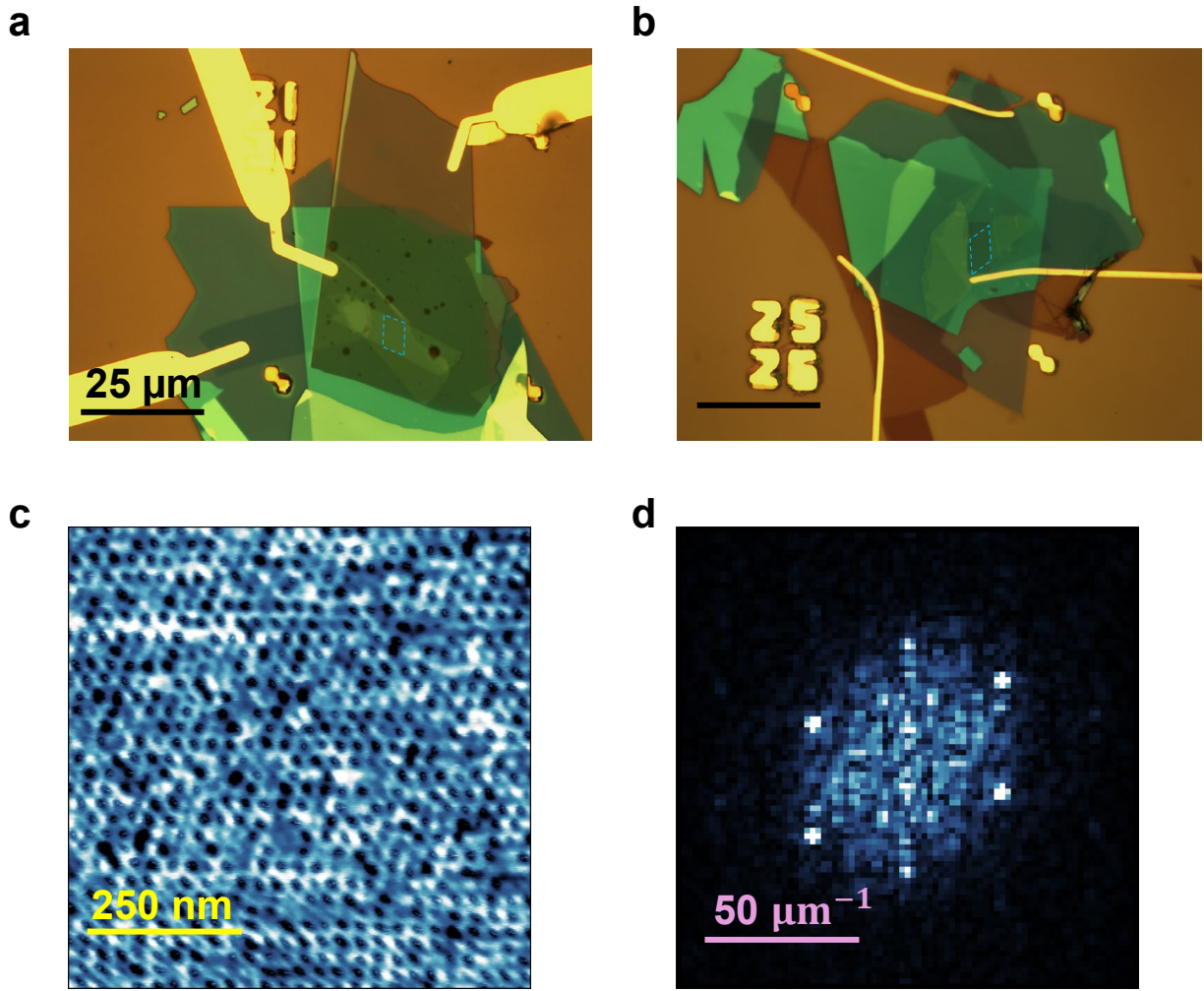

**Figure. S1. Device optical images and regularity of the patterned 30 nm lattice. a,** Optical micrograph of device A, an hBN encapsulated MoSe₂ heterostructure with a graphite bottom gate and a FLG top gate. Device A has a 40 nm periodic triangular lattice of 11 nm diameter holes etched into the top gate. The lattice perimeter is outlined in blue and extends over an area of approximately 28 square microns. **b,** Optical micrograph of the sample with a 30 nm triangular lattice. The lattice is a rectangle of approximately 27 square microns and is outlined in blue. **c,** AFM topographic image of the 30 nm lattice patterned into the PMMA mask before etching the FLG gate. Individual holes are 11 nm in diameter. **d,** Spatial Fourier transformation of the AFM

image showing six peaks associated with the lattice vectors. The lattice constant is determined to be $a = 28.67$ nm by fitting 2D Gaussian functions to the peaks. The full width at half maximum variability of the lattice constant is $\Delta a = 2.65$ nm.

## 2. Determination of charge density and sample homogeneity

Charge density was calculated for the patterned devices by simulating the capacitance of each structure, considering the appropriate etched-gate geometry. The charge density of the unpatterned device was calculated assuming a parallel-plate capacitor model for the gates. To account for variance in the hBN dielectric constants between devices, we scaled the charge density of devices B and C such that their exciton blue shift matched the blue shift of device A over a sufficiently large gate voltage range. The blue shift over a $n = 3.1 \times 10^{12}$ cm$^{-2}$ doping range is shown in Fig. S2a,b. This scaling ensures that all devices have a consistent definition of charge density. Finally, the charge density for all devices was scaled down by an additional 13% to account for variations in the dielectric constant of our hBN from the constant used in our simulations. We note that while the scaling of charge density in devices A and B can be accounted for by variations in $\epsilon_{hBN}$, in device C the scaling required to match exciton blue shifts is too large to be explained by variations[1] in the dielectric constant. While measuring, the bottom gate in device C was not functioning and left floating while the top gate was used to dope the sample. This can explain the large discrepancy between calculated charge density in this device when compared to devices A and B. Because device C is used as a control to demonstrate the lack of $\frac{dE}{dn}$ peaks, an exact calibration of charge density is not crucial.

The $\frac{dE}{dn}$ curve for the three devices after scaling the charge density in devices B and C is plotted in Fig. S2c. The first prominent exciton blue shift is marked by a red dot and was used to ascribe the value of $n = 0$. The displacement field applied during telegraph noise scans was calculated as $D/\epsilon_0 = \epsilon(\frac{V_t}{l_t} - \frac{V_b}{l_b})$, where $\epsilon_0$ is the permittivity of free space and $\epsilon$ is the dielectric constant of the surrounding hBN.

Figure S2d shows Fourier transformed $\frac{dE}{dn}$ curves for device A during different cooldown cycles and at different probing locations. The location of periodic peaks is consistent across the sample, showing strong structure at the relevant values of $P$. The $\frac{dE}{dn}$ peak amplitude varies across the sample giving rise to variations in the amplitude of the Fourier transformed curve, but the location remains consistent.

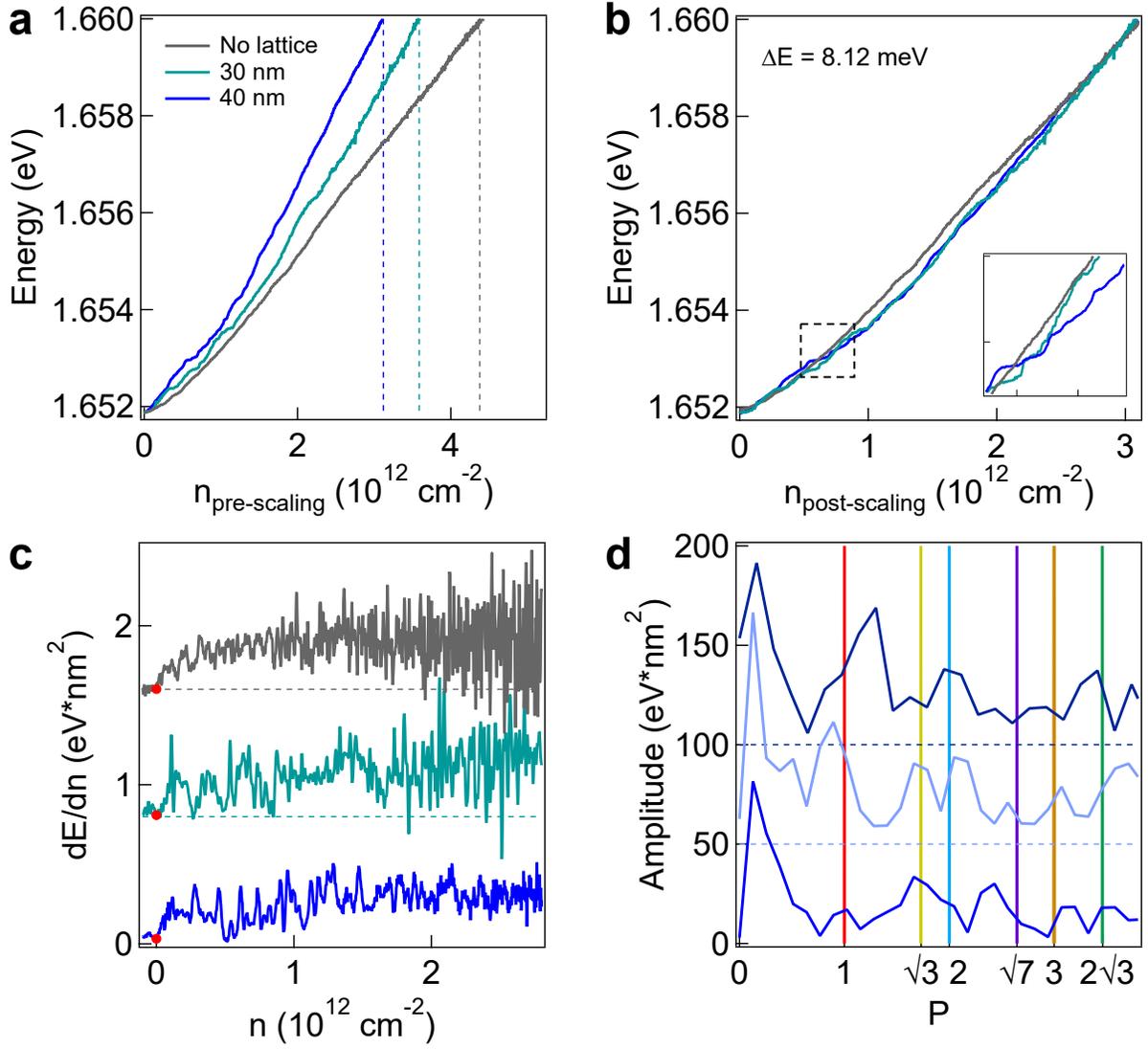

**Figure. S2. Determination of charge density and homogeneity across the sample. a,** Exciton energy in the three devices as a function of capacitively calculated doping. Corrections to the doping in devices B and C were made by matching their exciton blue shift to that of device A over a $3.1 \times 10^{12}$ cm$^{-2}$ doping range. The dashed lines indicate where $n = 3.1 \times 10^{12}$ cm$^{-2}$ was for device A, and where it was scaled to for devices B and C. **b,** Exciton energy as a function of doping after the scaling to devices B and C was applied. The total blue shift for each device is 8.12 meV over a $n = 3.1 \times 10^{12}$ cm$^{-2}$ doping range after scaling. Beside expected deviations at densities below $n = 2.5 \times 10^{12}$ cm$^{-2}$ due to emerging band gaps in the patterned samples, the blue shifting trend is consistent between the three devices. The inset highlights the jumps in the patterned samples. **c,** $\frac{dE}{dn}$ curves for the three samples as functions of charge density. The location at which $n = 0$ was chosen is marked by red dots and was determined by the first prominent blue shift. **d,** Fourier analysis of $\frac{dE}{dn}$ peaks in device A during different cooldown cycles and at different probing locations on the lattice. The location of peaks remains consistent across the sample.

# 3. Identification of specific Wigner crystal features

It is interesting to note that while the high density of peaks makes it difficult to identify Wigner crystals with discrete and specific densities, analyzing linecuts of $\frac{dE}{dn}$ along various fixed displacement fields allows us to rule out several peaks because they do not occur at fixed charge densities. Fig. S3a shows $\frac{dE}{dn}$ as a function of gate voltages in device A. Several peaks are annotated by short lines and denote features that are independent of field. As highlighted in Fig. S3b, the density that a given Wigner crystal peak occurs at is independent of displacement field. While several other optical features do appear to depend on displacement field, they cannot be attributed to Wigner crystals and are left as the subject of future study. Optical features in this extended gate voltage space is further explored in Fig. S3c-f where $\frac{dE}{dn}$ is plotted for T = 1.6 K, 15 K, 30 K, and 45 K respectively. At T = 1.6 K peaks occur at all displacement field values shown, however by T = 30 K only one peak is resolvable and occurs exclusively at finite field values. At T = 45 K all sharp $\frac{dE}{dn}$ are gone, signifying a smoothly blue shifting exciton. Many features persist to elevated temperatures when compared to continuous Wigner crystals[2] in MoSe$_2$. The addition of an external periodic potential increases the melting temperature due to pinning by potential sites that increases localization and reduces kinetic motion[3].

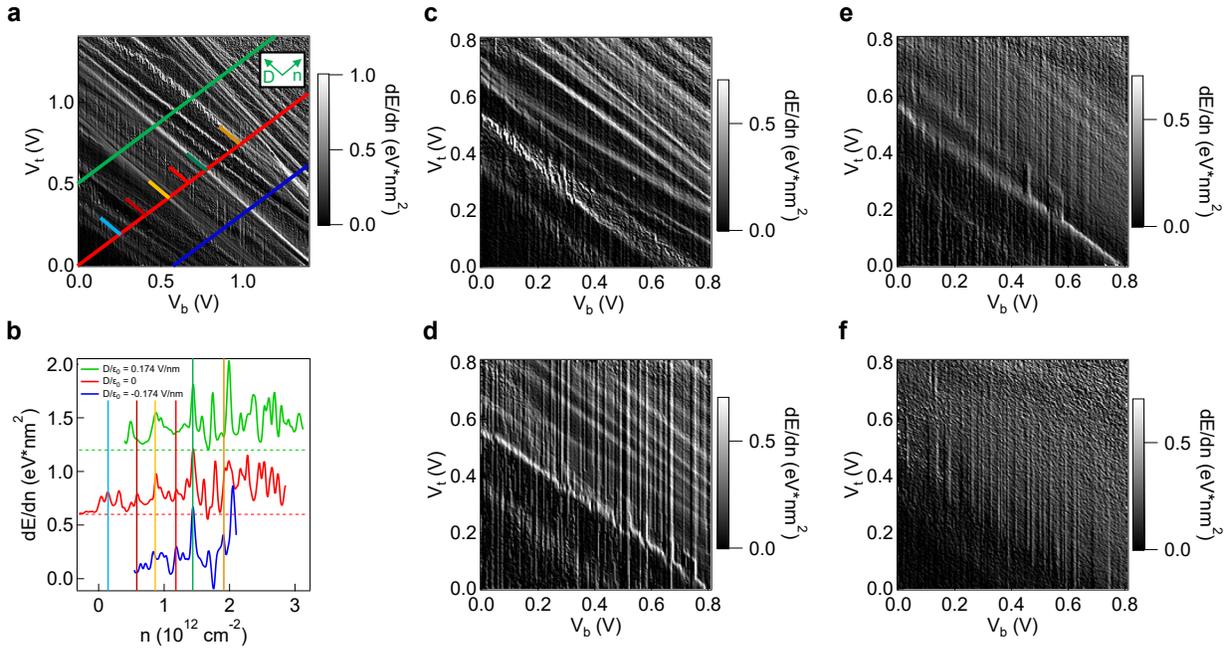

**Figure. S3. Field and temperature dependence of dE/dn spikes in device A. a,** Map of $\frac{dE}{dn}$ as a function of gate voltages. Three doping-dependent linecuts are labelled by green, red and blue lines. **b,** Linecuts of the $\frac{dE}{dn}$ plot in **(a)** at $\frac{D}{\varepsilon_0} = 0.174\,\frac{V}{nm}$, $\frac{D}{\varepsilon_0} = 0\,\frac{V}{nm}$, and $\frac{D}{\varepsilon_0} = -0.174\,\frac{V}{nm}$. The features annotated by vertical lines are independent of field and are attributed to Wigner crystals while other features dependent on field and are not attributable to Wigner crystals. **c,** $\frac{dE}{dn}$ Spectra as a function of gate voltages at T = 1.6 K. Several features, dependent and independent of displacement field, are present over the gate range shown. **d,** At T = 15 K many of the features persist while others diminish. Vertical lines are scanning artifacts. **e,** At T = 30 K most features are gone, with one feature existing only at finite displacement fields. **f,** At T = 45 K all sharp blue shifting exciton features are gone.

### 4. $k_f$-Periodic peaks at large values of *P*

Figure S4a shows the Fourier transformed $\frac{dE}{dn}$ curve over an extended range of *P* values for devices A, B and C. The two patterned samples decrease in amplitude with increasing *P* while the unpatterned device is dominated by noise up to *P* = 30. The reduced noise at high *P* values in the patterned samples is caused by data averaging on these devices over many scans. Data collected from the unpatterned sample is taken over a single scan and has larger noise as a result.

### 5. Temperature dependent Fourier transformation of exciton blueshifts

Figure S4b shows the temperature dependence of the exciton blue shifts with doping. As temperature increases, the blue shifting becomes smooth and the peaks are less defined. The Fourier transformation of these curves is plotted in Fig. S4c and shows diminishing intensity in periodic features up to *T* = 15 K where only the peaks periodic with *P* = 1 and *P* = 2 remain. The dashed gray lines indicate integer multiples of $P = \sqrt{3}$ and the dotted gray line indicates $P = \sqrt{7}$.

### 6. Fourier analysis near the hole-doped band edge

Figure S4d shows the doping dependence of the exciton near the hole-doped band edge. While the energy (red curve) blue shifts with increased hole doping, noise on the order of ~ 1 meV obscures any sharp blue shifting associated with Wigner crystallization. The $\frac{dE}{dn}$ curve associated with the hole-doped exciton energy is plotted in black. Noise in energy gives rise to peaks on the order of ~ 5 $\frac{eV}{nm^2}$. While Fourier analysis increases the signal-to-noise ratio for the electron-doped exciton energy, at the hole-doped edge the Fourier transformed $\frac{dE}{dn}$ curve remains noisy with no significant matching to Wigner crystal peaks as shown in Fig. S4e.

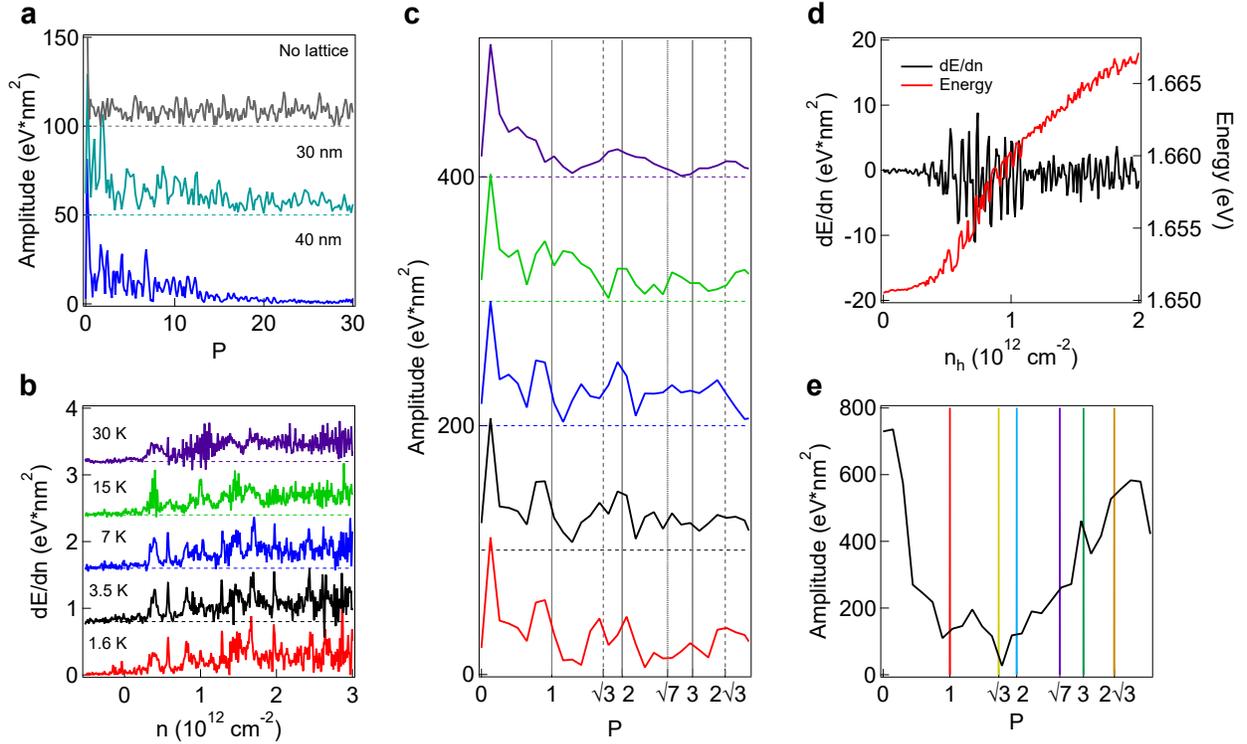

**Figure. S4. Extended range Fourier transforms and crystal melting. a,** Fourier transformed exciton energy for the three devices over a large range of $P$ values. The two patterned samples decrease in amplitude with increasing $P$ while the unpatterned device is dominated by noise up to $P = 30$. The reduced noise at high $P$ values in the patterned samples is caused by data averaging on these devices over many scans. Data collected from the unpatterned sample is taken over a single scan and has larger noise as a result. **b,** Temperature dependent blue shifting shows a steady decrease in the amplitude of all energy jumps up to $T = 30$ K where the sharp jumps are replaced by smooth blue shifting. **c,** Fourier transformed temperature-dependent blue shifts. The dashed gray lines indicate integer multiples of $P = \sqrt{3}$ and the dotted line shows $P = \sqrt{7}$. The $P = 1$ and $P = 2$ peaks remain until $T = 15$ K while the others disappear. **d,** Exciton energy and $\frac{dE}{dn}$ as a function of hole density at the hole-doped band edge. The energy shows fluctuations on the order of $1 - 2$ meV that cause peaks in $\frac{dE}{dn}$ that are ~ 10 times larger at the hole edge than the electron edge. **e,** Fourier transformation of the $\frac{dE}{dn}$ curve in **(d)** showing aperiodic noise with unresolvable structure at the relevant values of $P$.

## 7. References

(1) Hong, S.; Lee, C.-S.; Lee, M.-H.; Lee, Y.; Ma, K. Y.; Kim, G.; Yoon, S. I.; Ihm, K.; Kim, K.-J.; Shin, T. J.; Kim, S. W.; Jeon, E.-c.; Jeon, H.; Kim, J.-Y.; Lee, H.-I.; Lee, Z.; Antidormi, A.; Roche, S.; Chhowalla, M.; Shin, H.-J.; et al. Ultralow-Dielectric-Constant Amorphous Boron Nitride. *Nature* **2020**, *582* (7813), 511–514. DOI: 10.1038/s41586-020-2375-9.

(2) Smoleński, T.; Dolgirev, P. E.; Kuhlenkamp, C.; Popert, A.; Shimazaki, Y.; Back, P.; Lu, X.; Kroner, M.; Watanabe, K.; Taniguchi, T.; Esterlis, I.; Demler, E.; Imamoğlu, A. Signatures of Wigner Crystal of Electrons in a Monolayer Semiconductor. *Nature* **2021**, *595* (7865), 53–57. DOI: 10.1038/s41586-021-03590-4.

(3) Erkensten, D.; Brem, S.; Perea-Causin, R.; Malic, E. Stability of Wigner Crystals and Mott Insulators in Twisted Moiré Structures. *Physical Review B* **2024**, *110* (15), 155132. DOI: 10.1103/PhysRevB.110.155132.